\documentstyle[12pt]{article}
\normalbaselines
\begin{document}
\bibliographystyle{unsrt}
\def\beq{\begin{equation}}         \def\enq{\end{equation}}
\def\beqnar{\begin{eqnarray}}      \def\enqnar{\end{eqnarray}}
\def\pl{\partial}   \def\mat{\pmatrix}

\vbox{\vspace{6mm}} 

\begin{center}
{\large \bf CONFINEMENT IN RELATIVISTIC POTENTIAL MODELS}

\vspace{3cm}

J. Sucher\\ [5mm]
{\it Department of Physics, University of Maryland,
College Park, Maryland 20742}\\[5mm]

\end{center}
\vspace{2mm}

\begin{abstract}

     In relativistic potential models of quarkonia based on a
Dirac-type of equation with a local potential there is a sharp
distinction between a linear potential V which is vector-like and
one which is scalar-like: There are normalizable solutions for a
scalar-like V but not for a vector-like V.  It is pointed out that
if instead one uses an equation of the no-pair type, which is more
natural from the viewpoint of field theory, this somewhat bizarre
difference disappears.

\end{abstract}

\vspace{3cm}

1.  Since the discovery of the narrow resonances in the GeV region,
interest in a potential model description of these mesons and less
charming ones as quark-antiquark bound states has continued
unabated.  A long-standing problem which arises in this connection
is the following: If one tries to include relativistic effects with
a Dirac-type of equation involving a purely local potential there is
a dramatic difference between a linear potential which is vector-like,
 $V = k_{v}r$,  and one which is scalar-like, $V = \beta_1\beta_2 k_sr$.
   For the
vector-like case there are no normalizable solutions.  In view of
the continued interest in such models\cite{1,2}, it may be useful to point
out that if one uses equations of the no-pair type \cite{3}, which are much
more natural in the context of field theory, this dichotomy is not
present.

By way of review and for simplicity let us consider the case of an
antiquark much heavier than the quark and take as a starting point
a Dirac equation of the form
\beq
     (h_D + V)\psi = E\psi, \hspace{1cm} h_D = {\bf \alpha\cdot p}_{op} + \beta
m,
\enq
with $V$ linear in $r$.  A simple way to see the trouble which arises
for $V = k_vr$ is to decompose the wave function $\psi$ into a sum
\beq
      \psi = \psi^++ \psi^-, \hspace{1cm} \psi^\pm \equiv \beta^\pm \psi
\enq
where the $\beta^\pm$  are zero-momentum projection operators, defined by
\beq
     \beta^\pm \equiv (1\pm\beta)/2.
\enq
In the standard representation of the Dirac matrix $\beta$ this is
essentially a decomposition into upper and lower components, but no
use need be made of this fact.  From (1) and (2) we have
\beq
      \psi^- = (E+m-k_vr)^{-1} {\bf\alpha} \cdot {\bf p}_{op}
\enq
With $E > 0$, this implies that  $\psi^-$ has a pole singularity at $r =
(E+m)/k_v$ and is therefore not integrable.  Since
\beq
     <\psi|\psi> = <\psi^+|\psi^+> + <\psi^-|\psi^->,
\enq
the norm of $\psi^-$ will be infinite even if that of $\psi^+$ is finite.
However, if the potential is scalar-like, $V = \beta k_sr$, the minus sign
in the denominator in (4) changes to a plus sign,
\beq
     \psi^- = (E+m+k_sr)^{-1} {\bf \alpha\cdot p}_{op}\psi^+,
\enq
and there is no singularity.  Thus, if both scalar and vector
confining potentials are used it is necessary to have $k_s > k_v$.  The
same feature holds in the two-body equations of a similar type.\\

2. The corresponding no-pair equation does not suffer from this
dichotomy.  The counterpart of (1) is now
\beq
     (h_D + \Lambda_+^{op}U \Lambda_+^{op})\psi_+ = E\psi_+
\enq
where $\Lambda_+^{op}$ is the positive-energy Casimir projection operator,
defined by
\beq
  \Lambda_+^{op} = (E_{op}+h_D)/2E_{op},\ \  E_{op} \equiv ({\bf
p}_{op}^2+m^2)^{1/2}.
\enq
The subscript "+" indicates that $\psi_+$ satisfies
\beq
      \Lambda_+^{op}\psi_+ = \psi_+
\enq
and is thus a superposition of only positive-energy plane waves.
{}From (9) it follows that
\beq
      \psi_+^- = R_{op}\psi_+^+,~~~ R_{op} \equiv (E_{op}+m)^{-1} {\bf \alpha}
\cdot {\bf p}_{op}
\enq
Thus, regardless of the choice of $U$, there is now no singularity
involved in the equation relating  $\psi_+^+$ and $\psi_+^-$.  Furthermore, we
have
 $$<\psi_+^-|\psi_+^->\ =\ <\psi_+^+|R_{op}^{\dagger} R_{op}|\psi_+^+>$$
and in p-space the operator $R_{op}^{\dagger} R_{op}$
is just $p^2/[E(p)+m]^2$, which is bounded by unity. Thus
\beq
     <\psi_+^-|\psi_+^-> ~~<~~   <\psi_+^+|\psi_+^+>
\enq
and if the norm of $\psi_+^+$ is finite, so is that of $\psi_+^-$ and hence
that
of $\psi_+$.

3. To complete the argument let us compare the Schroedinger-Pauli
form of the eigenvalue problem for the two cases of interest.  As
in earlier work it is convenient to introduce a new wave function
$\phi$ which in p-space differs from $\psi_+^+$ by a slowly varying
factor \cite{3,4},
\beq
      \psi_+^+ = A_{op}\phi,~~~  A_{op} \equiv (E_{op}+m)/2E_{op}.
\enq
Then $\psi_+ = \psi_+^+ + \psi_+^- = (1+R_{op})\psi_+^+$, or
\beq
     \psi_+ = S\phi,~~~  S \equiv A_{op}(1+R_{op})\beta^+.
\enq
It is easy to verify that because of the extra factor $\beta^+$, $S$ is
pseudo-unitary, $S^{\dagger}S = \beta^+$. Since
\beq
      \beta^+\phi = \phi
\enq
it follows that $\psi_+$ and $\phi$ have the same norm.  On multiplying
(9) on the left by $S^{\dagger}$ one finds that $\phi$ satisfies the equation
\beq
     H_{red}\phi = E\phi
\enq
where
\beq
     H_{red} = "S^{\dagger}HS".
\enq
The quotes indicate that $\beta$  is to be replaced by unity when acting
directly on $\phi$.  For a potential $U$ of the generic form
\beq
     U = U_v + \beta U_s
\enq
computation yields
\beq
     H_{red} = E_{op} + V_{red},
\enq
where
\beq
     V_{red} = A_{op}U_+A_{op} +
     (2E_{op}A_{op})^{-1}{\bf \sigma\cdot p}_{op}U_-
     {\bf \sigma \cdot p}_{op}(2E_{op}A_{op})^{-1}
\enq
with
\beq
     U_{\pm} = U_v \pm U_s.
\enq
Since
 $${\bf \sigma\cdot p}_{op}U_-{\bf \sigma\cdot p}_{op} = {\bf p}_{op}\cdot U_
-{\bf p}_{op}
 +{\bf \sigma}\cdot(grad U_-)\times {\bf p}_{op},$$
  the main difference
between a pure vector and pure scalar potential is a change in sign
of part of the spin-independent relativistic correction and in the
sign of the spin-orbit interaction.  Since these corrections do not
dominate the effective interaction,  one expects that there are
normalizable solutions both in the scalar case and in the vector
case \cite{5}. \\

4. Of the making of potentials, as for books, there is no end.  One
criterion in a semi-phenomenological analysis of systems for which
it is makes sense to attempt a description in terms of relativistic
Schroedinger-like equations is simplicity.  The use of purely local
potentials lends itself to this because it limits the proliferation
of parameters.  Another criterion is to take note of the implications
of field theory.  For two spin-1/2 particles three-dimensional equations
tied to field theory inevitably lead to effective
interactions which involve projection operators.  A reasonable compromise
is therefore to consider equations of the no-pair type
\cite{3,4}:
\beq
     [(\alpha_1\cdot p_{op}+ \beta_1 m_1) +
     (-\alpha_2\cdot p_{op}+\beta_2m_2) + V_{++}]\psi  = E\psi
\enq
where
\beq
     V_{++} =  \Lambda_{++}U\Lambda_{++},
\enq
$\Lambda_{++}$ is the projection operator product  $\Lambda_+^{op}(1)
\Lambda_+^{op}(2)$ and
\beq
      \Lambda_+^{op}(i)\psi = \psi~~~(i = 1,2).
\enq
One may choose $U$ to be purely local without running into difficulties.
Note that the nonlocality of the projection operators does
not introduce any new parameters, since it involves only the
constituent masses, already present in the Dirac Hamiltonians.

As confirmation of the fact that no problems arise even if $U$ is
purely scalar-like, it should be noted that no difficulties are
encountered with the numerical solution of (21) when $U$ is chosen
to have the scalar form \cite{6}
\beq
     U = k\beta_1\beta_2r.
\enq
\vspace{2cm}
\begin{center}
                           Acknowledgements
\end{center}

     This work was supported in part by the National Science
Foundation. I thank Marshall Baker and Martin Olsson for raising
the issue addressed in this note.

\end{document}